\documentclass[aps,prl,showpacs,floatfix,twocolumn,byrevtex,superscriptaddress]{revtex4-1}
\usepackage{amsmath}
\usepackage{amssymb}
\usepackage{amstext}
\usepackage{amsopn}
\usepackage{amsfonts}
\usepackage{amsxtra}
\usepackage[english]{babel}
\usepackage{graphicx}
\usepackage{float}
\usepackage{bm}
\usepackage{multirow}
\usepackage{dcolumn}
\usepackage{color}
\usepackage{hyperref}

\newcommand{\icm}{\ensuremath{\textrm{cm}^{-1}}}% % cm-1

\newcommand{\KVS}{KV$_{3}$Sb$_{5}$}
\newcommand{\RVS}{RbV$_{3}$Sb$_{5}$}
\newcommand{\CVS}{CsV$_{3}$Sb$_{5}$}
\newcommand{\AVS}{$A$V$_{3}$Sb$_{5}$}
\newcommand{\EF}{$E_{\text{F}}$}
\newcommand{\TCDW}{$T_{\text{CDW}}$}

\begin{document}

\title{Electronic Correlations and Evolution of the Charge-Density Wave in the Kagome Metals $A$V$_{3}$Sb$_{5}$ ($A$ = K, Rb, Cs)}
\author{Xiaoxiang~Zhou}
\thanks{These authors contributed equally to this work.}
\affiliation{National Laboratory of Solid State Microstructures and Department of Physics, Collaborative Innovation Center of Advanced Microstructures, Nanjing University, Nanjing 210093, China}
\author{Yongkai~Li}
\thanks{These authors contributed equally to this work.}
\affiliation{Key Laboratory of Advanced Optoelectronic Quantum Architecture and Measurement, Ministry of Education, School of Physics, Beijing Institute of Technology, Beijing 100081, China}
\affiliation{Micronano Center, Beijing Key Lab of Nanophotonics and Ultrafine Optoelectronic Systems, Beijing Institute of Technology, Beijing 100081, China}
\author{Xinwei~Fan}
\thanks{These authors contributed equally to this work.}
\author{Jiahao~Hao}
\author{Ying~Xiang}
\author{Zhe~Liu}
\author{Yaomin~Dai}
\email{ymdai@nju.edu.cn}
\affiliation{National Laboratory of Solid State Microstructures and Department of Physics, Collaborative Innovation Center of Advanced Microstructures, Nanjing University, Nanjing 210093, China}
\author{Zhiwei~Wang}
\email{zhiweiwang@bit.edu.cn}
\author{Yugui~Yao}
\affiliation{Key Laboratory of Advanced Optoelectronic Quantum Architecture and Measurement, Ministry of Education, School of Physics, Beijing Institute of Technology, Beijing 100081, China}
\affiliation{Micronano Center, Beijing Key Lab of Nanophotonics and Ultrafine Optoelectronic Systems, Beijing Institute of Technology, Beijing 100081, China}
\author{Hai-Hu~Wen}
\email{hhwen@nju.edu.cn}
\affiliation{National Laboratory of Solid State Microstructures and Department of Physics, Collaborative Innovation Center of Advanced Microstructures, Nanjing University, Nanjing 210093, China}

\date{\today}
%%%%%%%%%%%%%%%%%%%%%%%%%%%%%%%%%%%%
%
% Abstract
%

\begin{abstract}
The kagome metals $A$V$_{3}$Sb$_{5}$ ($A$ = K, Rb, Cs) have attracted enormous interest as they exhibit intertwined charge-density wave (CDW) and superconductivity. The alkali-metal dependence of these characteristics contains pivotal information about the CDW and its interplay with superconductivity. Here, we report optical studies of $A$V$_{3}$Sb$_{5}$ across the whole family. With increasing alkali-metal atom radius from K to Cs, the CDW gap increases monotonically, whereas $T_{\text{CDW}}$ first rises and then drops, at variance with conventional CDW. While the Fermi surface gapped by the CDW grows, $T_{c}$ is elevated in CsV$_{3}$Sb$_{5}$, indicating that the interplay between the CDW and superconductivity is not simply a competition for the density of states near \EF. More importantly, we observe an enhancement of electronic correlations in CsV$_{3}$Sb$_{5}$, which suppresses the CDW but enhances superconductivity, thus accounting for the above peculiar observations. Our results suggest electronic correlations as an important factor in manipulating the CDW and its entanglement with superconductivity in $A$V$_{3}$Sb$_{5}$.
\end{abstract}

%  71.55.Ak  Metals, semimetals, and alloys
%  72.15.Eb  Electrical and thermal conduction in crystalline
%  78.20.-e  Optical properties of bulk materials and thin films
%  78.30.-j  Infrared and Raman spectra

%\pacs{78.20.-e, 78.30.-j}

\maketitle

%%%%%%%%%%%%%%%%%%%%%%%%%%%%%%%%%%%%%%%%%%%%%%%%%%%%%%%%%%%%%%%%%%%%%%%%%%%%%%
%
% Introduction
%
The kagome lattice, composed of hexagons and corner-sharing triangles, provides a fascinating playground for exploring exotic quantum phenomena. For instance, spins or magnetic moments on a kagome lattice are subject to high degree of geometric frustration that may lead to quantum spin liquids~\cite{Balents2010Nature,Yan2011Science}; electrons in a kagome lattice form flat bands, Dirac points, and saddle points, which support intriguing quantum phenomena associated with nontrivial band topology~\cite{Mazin2014NC,Ye2018Nature,Kang2020NM,Kang2020NC,Liu2020NC} and a wide variety of electronic instabilities~\cite{Yu2012PRB,Kiesel2012PRB,Kiesel2013PRL,Wang2013PRB}. Particularly at van Hove filling, as a function of the on-site repulsion $U$ and nearest-neighbor Coulomb interaction $V$, the kagome lattice exhibits a rich phase diagram consisting of various phases such as charge or spin bond order~\cite{Wang2013PRB,Kiesel2013PRL}, unconventional superconductivity~\cite{Yu2012PRB,Wang2013PRB,Kiesel2013PRL}, charge-density wave (CDW)~\cite{Wang2013PRB}, and spin-density wave (SDW)~\cite{Yu2012PRB}.

The recently discovered kagome metals \AVS\ ($A$ = K, Rb, Cs)~\cite{Ortiz2019PRM} with the Fermi level \EF\ lying near the saddle points (van Hove filling) provide an excellent platform to realize the above exotic quantum states in real materials. Multiple topologically protected Dirac bands~\cite{Ortiz2020PRL,Yin2021CPL} and superconductivity with a transition temperature $T_{c}$ of 0.92--2.5~K~\cite{Ortiz2020PRL,Ortiz2021PRM,Yin2021CPL} have been reported in these compounds. In addition, a CDW transition occurs at \TCDW\ = 78, 103, and 94~K for \KVS, \RVS, and \CVS, respectively~\cite{Ortiz2019PRM,Ortiz2020PRL,Ortiz2021PRM,Yin2021CPL}, resulting in a three-dimensional (3D) 2$\times$2$\times$2 superlattice~\cite{Liang2021PRX,Li2021PRX}. Upon entering the CDW state, a giant anomalous Hall effect~\cite{Yang2020SA,Yu2021PRB} and electronic nematicity~\cite{Xiang2021NC,Chen2021Nature,Li2022NP,Wu2021arXiv} also emerge. The application of pressure~\cite{Yu2021NC,Chen2021PRL,Du2021PRB,Zhang2021PRB}, uniaxial strain~\cite{Qian2021PRB}, or chemical doping~\cite{Song2021PRL,Oey2022PRM,Li2022PRBNb,Liu2021arXiv} suppresses the CDW order, but enhances the superconductivity, signifying the competition between the CDW and superconductivity in \AVS. While a variety of studies suggest that the saddle point or Fermi surface (FS) nesting plays an important role in driving the CDW instability~\cite{Tan2021PRL,Denner2021PRL,Zhou2021PRB,Christensen2021PRB,Wang2021arXiv,Cho2021PRL,Lou2022PRL,Rice1975PRL}, there is also evidence that the CDW phase is mainly driven by electron-phonon (e-ph) coupling~\cite{Luo2022NC,Xie2022PRB,Si2022PRB,Liu2022NC}. At the present time, the driving mechanism of the CDW and how it interacts with superconductivity in \AVS\ are subjects for intensive debate.

The evolution of the CDW and superconducting properties with alkali metal in \AVS\ may reveal key information about the factors controlling the CDW order and its interplay with superconductivity. In this Letter, we systematically study the optical properties of $A$V$_{3}$Sb$_{5}$ across the whole family. As the alkali-metal atom radius grows from K to Cs, the CDW gap $\Delta_{\text{CDW}}$ increases monotonically, whereas $T_{\text{CDW}}$ first rises but then drops, failing to exhibit a scaling relation with $\Delta_{\text{CDW}}$. This anomalous behavior is clearly at odds with the description of conventional CDW order. While the FS removed by $\Delta_{\text{CDW}}$ increases, $T_{c}$ is raised in \CVS, indicating that the interplay between the CDW and superconductivity is not simply a competition for the density of states (DOS) near \EF. Moreover, we observe an enhancement of electronic correlations in \CVS, which suppresses the CDW but promotes unconventional superconductivity, giving rise to the above peculiar behavior. Our results underline the importance of electronic correlations in manipulating the CDW and its entanglement with superconductivity in \AVS.

%%%%%%%%%%%%%%%%%%%%%%%%%%%%%%%%%%%%%%%%%%%%%%%%%%%%%%%%%%%%%%%%%%%%%%%%%%%%%%
%
% Experiment
%

%%%%%%%%%%%%%%%%%%%%%%%%%%%%%%%%%%
% Figure 1
\begin{figure}[tb]
\includegraphics[width=\columnwidth]{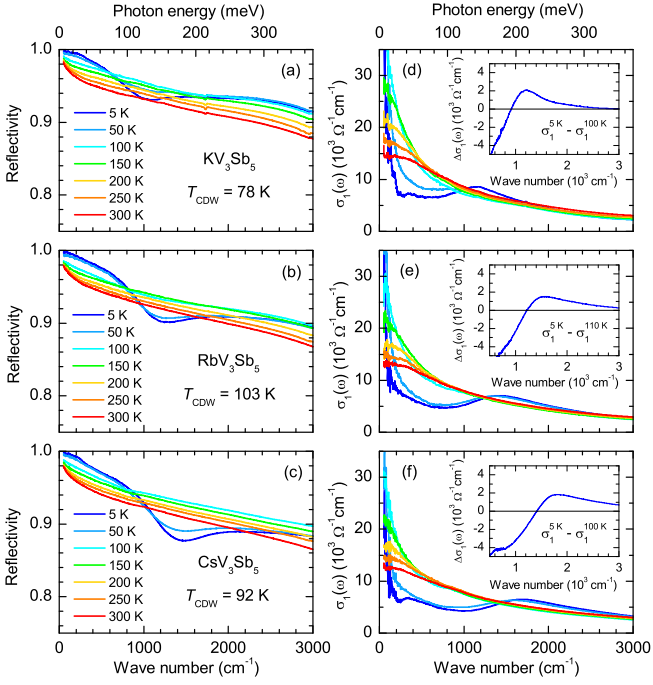}
\caption{(a)--(c) $R(\omega)$ at several representative temperatures for \KVS, \RVS, and \CVS, respectively. (d)--(f) $\sigma_{1}(\omega)$ for \KVS, \RVS, and \CVS, respectively. The inset displays $\Delta\sigma_{1}(\omega)$ at 5~K for each compound, where $\sigma_{1}(\omega)$ at $T$ just above $T_{\text{CDW}}$ is used as the base curve.}
\label{ARefS1}
\end{figure}

High-quality single crystals of \AVS\ ($A$ = K, Rb, Cs) were synthesized using the self-flux method and characterized by x-ray diffraction~\cite{SuppMat}. Transport measurements were carried out to confirm the CDW transition at \TCDW\ = 78, 103, and 92~K in \KVS, \RVS, and \CVS, respectively~\cite{SuppMat}. The details of optical measurements can be found in the Supplemental Material~\cite{SuppMat}.

%%%%%%%%%%%%%%%%%%%%%%%%%%%%%%%%%%%%%%%%%%%%%%%%%%%%%%%%%%%%%%%%%%%%%%%%%%%%%%
%
% Data analysis
%

Figures~\ref{ARefS1}(a)--\ref{ARefS1}(c) show the measured reflectivity $R(\omega)$ up to 3000~\icm\ at several representative temperatures for \KVS, \RVS, and \CVS, respectively. Above \TCDW, $R(\omega)$ for all three compounds exhibits metallic behavior: a very high $R(\omega)$ in the far-infrared range that approaches unity in the zero-frequency limit and increases with decreasing $T$. Below \TCDW, a suppression of $R(\omega)$ in the frequency range of 1000--1500~\icm\ occurs for all three materials, signaling the opening of a CDW gap. As the radius of the alkali-metal atom grows from K to Cs, the suppression in $R(\omega)$ shifts to higher frequency and deepens, suggesting that the CDW gap increases in energy and the gap-induced FS modification intensifies with growing alkali-metal atom radius.

%%%%%%%%%%%%%%%%%%%%
% Figure 2
\begin{figure}[tb]
\includegraphics[width=\columnwidth]{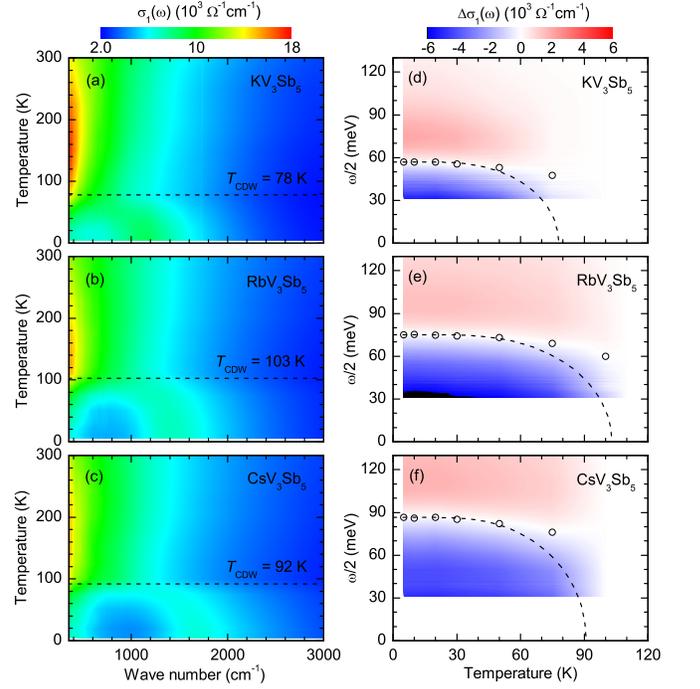}
\caption{(a)--(c) The 2D $T$-$\omega$ map of $\sigma_{1}(\omega)$ for \KVS, \RVS, and \CVS, respectively. The horizontal dashed line in each panel denotes $T_{\text{CDW}}$. (d)--(f) $\Delta\sigma_{1}(\omega)$ in the 2D $\omega/2$-$T$ plot, where the white color (zero crossing points) yields the $T$ dependence of $\Delta_{\text{CDW}}$. The dashed line in each panel represents the mean-field behavior.}
\label{AS1CM}
\end{figure}
Figures~\ref{ARefS1}(d)--\ref{ARefS1}(f) show the optical conductivity $\sigma_{1}(\omega)$ at several representative temperatures above and below \TCDW\ for \KVS, \RVS, and \CVS, respectively. For all three compounds, above $T_{\text{CDW}}$, a Drude response, i.e. a peak centered at zero frequency, can be clearly observed in the low-frequency $\sigma_{1}(\omega)$, in good agreement with the metallic nature of these materials; below \TCDW, a dramatic suppression of the low-frequency $\sigma_{1}(\omega)$ sets in, and the removed spectral weight [the area under $\sigma_{1}(\omega)$] is transferred to higher frequencies, which is the prototypical response of the CDW gap in $\sigma_{1}(\omega)$. The detailed evolution of $\sigma_{1}(\omega)$ with $T$ is traced out in the 2D temperature-frequency ($T$-$\omega$) maps in Figs.~\ref{AS1CM}(a)--\ref{AS1CM}(c) for all three materials. The horizontal dashed line in each panel denotes \TCDW\ for \KVS, \RVS, and \CVS, respectively. Below \TCDW, the opening of the CDW gap leads to the presence of a blue region in the low-frequency range [a suppression of the low-frequency $\sigma_{1}(\omega)$] and a shift of the green/cyan region to higher frequencies. A comparison of Figs.~\ref{AS1CM}(a)--\ref{AS1CM}(c) reveals that as the radius of the alkali-metal atom in \AVS\ increases, the low-frequency blue region moves to higher frequencies and grows in area. These observations indicate that not only does the CDW gap value increase, the removed spectral weight due to the opening of the CDW gap also grows with increasing alkali-metal atom radius.

\begin{figure}[tb]
\includegraphics[width=\columnwidth]{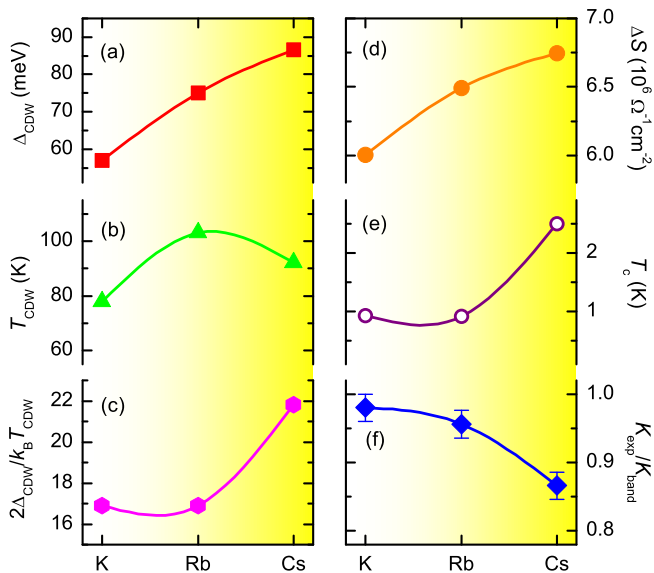}
\caption{The evolution of (a) $\Delta_{\text{CDW}}$, (b) $T_{\text{CDW}}$, (c) $2\Delta_{\text{CDW}}/k_{\text{B}}T_{\text{CDW}}$, (d) $\Delta S$, (e) $T_{c}$, and (f) $K_{\text{exp}}/K_{\text{band}}$ with the alkali-metal atom in the \AVS\ family.}
\label{AParvsA}
\end{figure}

The CDW gap $\Delta_{\text{CDW}}$ can be determined from the zero-crossing point in the difference optical conductivity,
%%%%%%%%%%%%%
%
% Eq.1
%
\begin{equation}
\Delta \sigma_{1}(\omega) = \sigma_{1}^{T < T_{\text{CDW}}}(\omega) - \sigma_{1}^{N}(\omega),
\end{equation}
where $\sigma_{1}^{T < T_{\text{CDW}}}(\omega)$ represents $\sigma_{1}(\omega)$ at $T < T_{\text{CDW}}$; $\sigma_{1}^{N}(\omega)$ refers to $\sigma_{1}(\omega)$ in the normal state, namely at $T$ slightly above \TCDW. The insets of Figs.~\ref{ARefS1}(d)--\ref{ARefS1}(f) display $\Delta\sigma_{1}(\omega)$ at 5~K for \KVS, \RVS, and \CVS, respectively, in which the zero-crossing point corresponds to $2\Delta_{\text{CDW}}$. Therefore, the $T$ dependence of $\Delta_{\text{CDW}}$ can be obtained by plotting $\Delta\sigma_{1}(\omega)$ in the 2D $\omega/2$-$T$ maps as shown in Figs.~\ref{AS1CM}(d)--\ref{AS1CM}(f) for \KVS, \RVS, and \CVS, respectively. In each panel, while the red and blue colors denote positive and negative values for $\Delta\sigma_{1}(\omega)$, the white color corresponds to the zero-crossing points which yields the evolution of $\Delta_{\text{CDW}}$ with $T$. For all three materials, the $T$ dependence of $\Delta_{\text{CDW}}$ deviates from the BCS mean-field behavior (black dashed line) in the proximity of \TCDW, consistent with previous studies~\cite{Zhou2021PRB,Uykur2021PRB,Uykur2022NPJQM}. This implies that the CDW transition in \AVS\ is unconventional and most likely to be of the first order, in accord with recent NMR studies~\cite{Mu2021CPL,Song2022SCPMA,Luo2022NPJQM}. Moreover, $\Delta_{\text{CDW}}$ increases monotonically with increasing alkali-metal atom radius, failing to show a scaling relation with \TCDW\ which increases in \RVS\ but then drops in \CVS. The values of $\Delta_{\text{CDW}}$ and \TCDW\ for \AVS\ are summarized in Figs.~\ref{AParvsA}(a) and \ref{AParvsA}(b), respectively, for further discussions.

The removed spectral weight $\Delta S$ in the low-frequency range due to the opening of $\Delta_{\text{CDW}}$ reflects the gapped portion of the FS or the reduction of the density of states (DOS) near \EF. $\Delta S$ can be directly obtained from the integral of $\sigma_{1}(\omega)$,
%%%%%%%%%%%%%
%
% Eq.2
%
\begin{equation}
\Delta S = \int_{0}^{2\Delta_{\text{CDW}}}\left[\sigma_{1}^{N}(\omega)-\sigma_{1}^{T < T_{\text{CDW}}}(\omega)\right]\text{d}\omega,
\label{DeltaS}
\end{equation}
The values of $\Delta S$ at $T$ = 5~K determined from Eq.~(\ref{DeltaS}) are 6.00, 6.49, and 6.74$\times 10^{6}$~$\Omega^{-1}$cm$^{-2}$ for \KVS, \RVS, and \CVS, respectively. The increase of $\Delta S$ with increasing alkali-metal atom radius indicates that a larger portion of the FS is removed by $\Delta_{\text{CDW}}$ in \AVS\ with a larger alkali-metal atom. We plot $\Delta S$ at $T$ = 5~K as a function of alkali metal in Fig.~\ref{AParvsA}(d) for further discussions.

%%%%%%%%%%%%%%%%%%%%
% Figure 3
\begin{figure}[tb]
\includegraphics[width=\columnwidth]{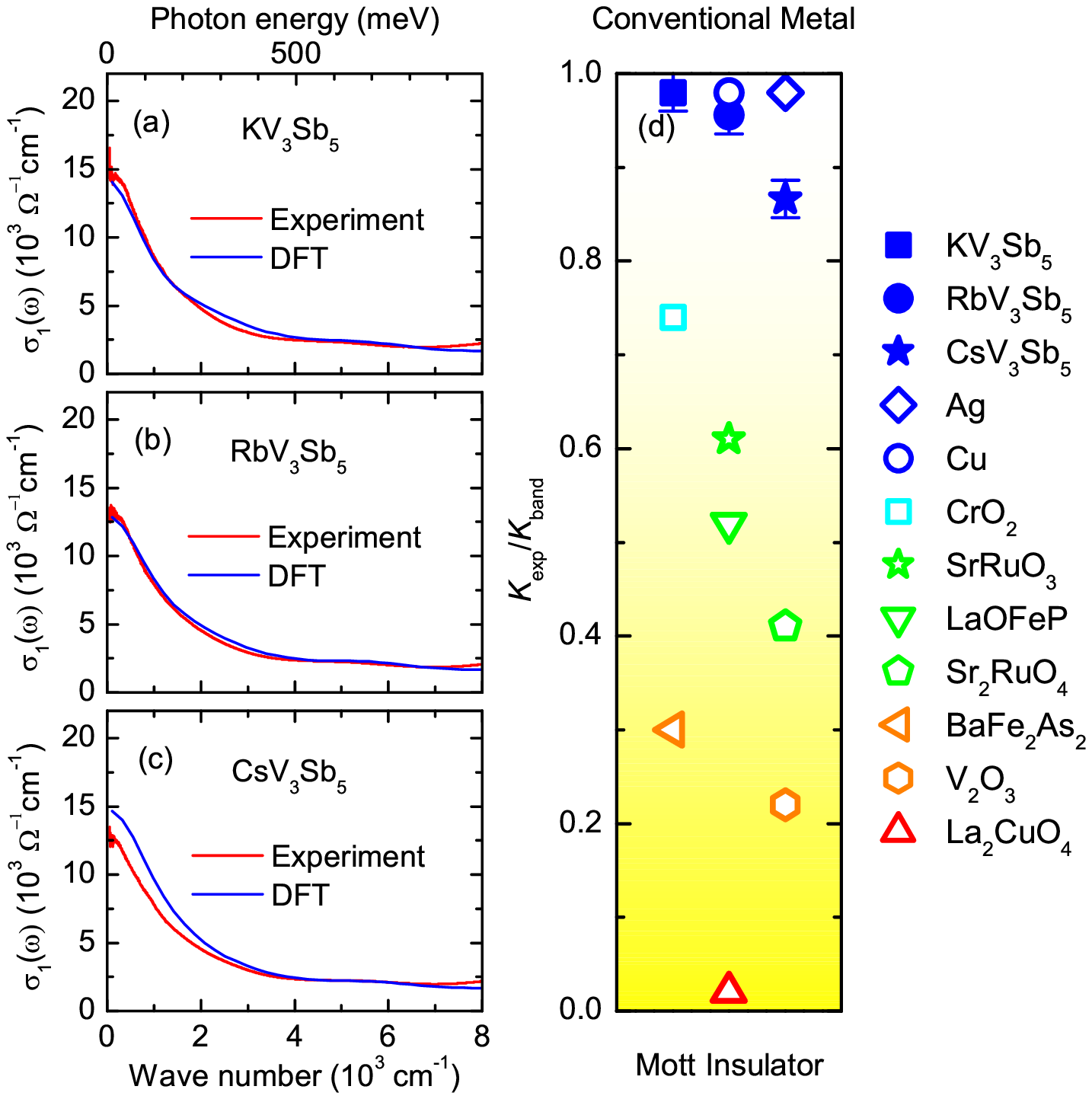}
\caption{(a)--(c) Comparison between the calculated $\sigma_{1}(\omega)$ (blue solid curve) and the measured $\sigma_{1}(\omega)$ at 300~K (red solid curve) for \KVS, \RVS, and \CVS, respectively. (d) $K_{\text{exp}}/K_{\text{band}}$ for \AVS\ (solid symbols) and several other representative materials (open symbols). The values of $K_{\text{exp}}/K_{\text{band}}$ for other materials are obtained from Ref.~\cite{Qazilbash2009NP} and the references cited therein.}
\label{ACor}
\end{figure}
The ratio of the experimental kinetic energy $K_{\text{exp}}$ to the kinetic energy from band theory $K_{\text{band}}$ provides crucial information about the electronic correlations in \AVS~\cite{Millis2005PRB,Qazilbash2009NP,Xu2020NC}. The electron's kinetic energy can be conveniently derived from the area under the Drude feature in $\sigma_{1}(\omega)$~\cite{Millis2005PRB,Qazilbash2009NP},
%%%%%%%%%%%%%
%
% Eq.4
%
\begin{equation}
K = \frac{2\hbar^{2} c_{0}}{\pi e^2}\int_{0}^{\omega_{c}}\sigma_{1}(\omega)\text{d}\omega,
\label{Kinetic}
\end{equation}
where $c_{0}$ is the distance between the V kagome layers, and $\omega_{c}$ is a cutoff frequency covering the entire Drude component in $\sigma_{1}(\omega)$. In order to determine $K_{\text{band}}$, we calculated the $ab$-plane $\sigma_{1}(\omega)$ for all three materials~\cite{SuppMat}. As depicted in Figs.~\ref{ACor}(a)--\ref{ACor}(c), the calculated $\sigma_{1}(\omega)$ spectra (blue curves) qualitatively agree with the measured ones (red curves). Using $\omega_{c}$ = 5000~\icm\ for both the measured and the calculated $\sigma_{1}(\omega)$, the values of $K_{\text{exp}}/K_{\text{band}}$ are obtained for all three compounds. Figure~\ref{ACor}(d) summarizes $K_{\text{exp}}/K_{\text{band}}$ for \AVS\ (solid symbols) and some other representative materials (open symbols). While conventional metals, such as Ag and Cu, have $K_{\text{exp}}/K_{\text{band}}$ close to unity, $K_{\text{exp}}/K_{\text{band}}$ for the well-known Mott insulator, such as La$_{2}$CuO$_{4}$, is almost zero due to on-site Coulomb repulsion which impedes the motion of electrons. Iron pnictides, e.g. LaOFeP and BaFe$_{2}$As$_{2}$, are categorized as moderately correlated materials, as their $K_{\text{exp}}/K_{\text{band}}$ lies between conventional metals and Mott insulators. The values of $K_{\text{exp}}/K_{\text{band}}$ for \KVS\ (solid square), \RVS\ (solid circle), and \CVS\ (solid star) fall into the range of 0.86--0.98, indicating that the electronic correlations in \AVS\ is weak in general. However, it's worth noting that while the values of $K_{\text{exp}}/K_{\text{band}}$ in \KVS\ and \RVS\ are close to that in conventional metals, \CVS\ features a reduced $K_{\text{exp}}/K_{\text{band}}$, signifying an enhancement of electronic correlations. Theoretical calculations have revealed that the FS of \AVS\ is formed by Sb-5$p$ and V-3$d$ orbitals~\cite{Tan2021PRL,Zhao2021PRB,LaBollita2021PRB}. Since 3$d$ electrons are subject to strong electronic correlations~\cite{Imada1998RMP,Yu2021FP}, the electronic correlations in \AVS\ are likely to arise from the V-3$d$ orbitals. Here, both $K_{\text{exp}}$ and $K_{\text{band}}$ consist of contributions from all orbitals at \EF, including the weakly correlated Sb-5$p$ orbital, so the actual electronic correlations in the V-3$d$ orbitals are stronger than that determined from $K_{\text{exp}}/K_{\text{band}}$. Given that the ground state of the kagome lattice sensitively depends on electronic correlations~\cite{Wang2013PRB,Kiesel2013PRL,Denner2021PRL}, the enhancement of electronic correlations in \CVS, although not as strong as that in cuprates and iron pnictides, may have an important influence on the CDW and its interplay with superconductivity. For further discussions, the evolution of $K_{\text{exp}}/K_{\text{band}}$ with alkali metal is traced out in Fig.~\ref{AParvsA}(f).

%%%%%%%%%%%%%%%%%%%%
% Figure 4
%

%%%%%%%%%%%%%%%%%%
% Discussion
%

Essential information about the CDW and how it intertwines with superconductivity in \AVS\ can be obtained from the alkali-metal dependence of the experimentally determined parameters, which has been summarized in Figs.~\ref{AParvsA}(a)--\ref{AParvsA}(f). A noticeable yet puzzling observation is that while $\Delta_{\text{CDW}}$ [Fig.~\ref{AParvsA}(a)] increases monotonically with increasing alkali-metal atom radius, \TCDW\ [Fig.~\ref{AParvsA}(b)] first increases but then decreases, failing to establish a scaling relation with $\Delta_{\text{CDW}}$. This anomalous behavior is clearly at odds with the conventional description of CDW order~\cite{Gruner1988RMP}. Figure~\ref{AParvsA}(c) shows that \KVS\ and \RVS\ have the same $2\Delta_{\text{CDW}}/k_{\text{B}}T_{\text{CDW}}$, indicating that the CDW is governed by the same mechanism in these two compounds, whereas $2\Delta_{\text{CDW}}/k_{\text{B}}T_{\text{CDW}}$ is significantly larger in \CVS, suggesting that an extra factor is acting on the CDW order. Thus far, the driving force of the CDW order in \AVS\ is still highly controversial. Extensive studies hint that the CDW instability in \AVS\ is driven by the nesting of the FS or saddle points~\cite{Tan2021PRL,Denner2021PRL,Christensen2021PRB,Zhou2021PRB,Wang2021arXiv,Cho2021PRL,Lou2022PRL,Rice1975PRL}, but on the other hand, a recent ARPES study on \KVS~\cite{Luo2022NC}, neutron scattering~\cite{Xie2022PRB} and Raman~\cite{Liu2022NC} measurements on \CVS\ suggest that e-ph coupling plays a dominant role in driving the CDW instability in \AVS. Nevertheless, for both the nesting driven and e-ph coupling driven scenarios, a larger $\Delta_{\text{CDW}}$ would naturally support a higher \TCDW. Here in \CVS, an increase of $\Delta_{\text{CDW}}$ coincides with a decrease of \TCDW, implying that besides e-ph coupling and FS nesting which should enhance both $\Delta_{\text{CDW}}$ and \TCDW, a competing factor that suppresses \TCDW\ also exerts considerable influence on the CDW order. Previous studies on transition metal dichalcogenides have documented that while e-ph coupling, FS nesting, and a high DOS are beneficial to the formation of CDW order, electronic correlations act as a competing factor which tends to localize the carriers and prevents the formation of CDW order~\cite{Loon2018NPJQM,Lin2020NC}. Our optical results have attested to the decrease of $K_{\text{exp}}/K_{\text{band}}$ [Fig.~\ref{AParvsA}(f)], i.e. the enhancement of electronic correlations in \CVS. These facts bring us to the possibility that the suppression of $T_{\text{CDW}}$ and the larger $2\Delta_{\text{CDW}}/k_{\text{B}}T_{\text{CDW}}$ in \CVS\ may be related to the enhancement of electronic correlations.

Other interesting observations emerge from the comparison of $T_{c}$, \TCDW, $\Delta S$, and $2\Delta_{\text{CDW}}/k_{\text{B}}T_{\text{CDW}}$. As shown in Figs.~\ref{AParvsA}(b) and \ref{AParvsA}(e), the suppression of \TCDW\ in \CVS\ is accompanied by an enhancement of $T_{c}$, in accord with the competition relation between the CDW and superconductivity reported by previous work~\cite{Yu2021NC,Chen2021PRL,Du2021PRB,Zhang2021PRB,Qian2021PRB,Song2021PRL,Oey2022PRM}. The competition between the CDW and superconductivity is not surprising, because the opening of the CDW gap depletes the DOS near \EF, resulting in a suppression of superconductivity. However, the comparison of Figs.~\ref{AParvsA}(d) and \ref{AParvsA}(e) reveals that the rise of $T_{c}$ in \CVS\ coincides with an increase of $\Delta S$. This observation implies that the CDW instability and superconductivity in \AVS\ do not share the same DOS, so that the competition between the CDW and superconductivity is not simply through competing for effective DOS near \EF, but controlled by another factor. Furthermore, $2\Delta_{\text{CDW}}/k_{\text{B}}T_{\text{CDW}}$ [Fig.~\ref{AParvsA}(c)] and $T_{c}$ [Fig.~\ref{AParvsA}(e)] exhibit identical alkali-metal dependence, hinting that the suppression of \TCDW\ and the enhancement of $T_{c}$ in \CVS\ are most likely induced by the same factor, namely electronic correlations. A recent theoretical study has shown that electronic correlations suppress the charge susceptibility but significantly enhance the spin susceptibility or spin fluctuations~\cite{Loon2018NPJQM}, which are believed to mediate unconventional superconductivity~\cite{Moriya2000AP}. Moreover, extensive studies on cuprates and iron pnictides have underlined the importance of electronic correlations in generating unconventional superconductivity~\cite{Lee2006RMP,Yu2021FP}. Combining these studies with our observations, we suggest that electronic correlations should be taken into account when constructing a theory to describe the CDW phase and its entanglement with superconductivity in \AVS.

%%%%%%%%%%%%%%%%%%%%%%%%%%%%%%%%%%%%%%%%%%%%%%%%%%%%%%%%%%%%%%%%%%%%%%%%%%%%%%
%
% Conclusions
%
To summarize, we performed a systematic investigation into the optical properties of $A$V$_{3}$Sb$_{5}$ ($A$ = K, Rb, Cs) across the whole family. We found that as the alkali-metal atom radius grows from K to Cs, (i) while $\Delta_{\text{CDW}}$ increases monotonically, $T_{\text{CDW}}$ rises in \RVS\ but then drops in \CVS, at odds with conventional CDW; (ii) the FS removed by $\Delta_{\text{CDW}}$ increases, whereas $T_{c}$ is enhanced in \CVS, suggesting that the interplay between the CDW and superconductivity is not simply a competition for effective DOS near \EF; (iii) $K_{\text{exp}}/K_{\text{band}}$ is reduced, indicating an enhancement of electronic correlations. An analysis considering all the above observations and previous work suggests that the enhancement of electronic correlations may be a decisive factor that controls the formation of the CDW order and its entanglement with superconductivity in \AVS.

%%%%%%%%%%%%%%%%%%%%%%%%%%%%%%%%%%%%%%%%%%%%%%%%%%%%%%%%%%%%%%%%%%%%%%%%%%%%%%
%
% Acknowledgment
%

\begin{acknowledgments}
We thank Hu~Miao, R.~Thomale, Qianghua~Wang, Nanlin~Wang, Xiaoxiang~Xi, Bing~Xu, Binghai~Yan, Huan~Yang, Run~Yang, Shunli~Yu, Peng~Zhang, and Jianzhou Zhao for helpful discussions. We gratefully acknowledge financial support from the National Key R\&D Program of China (Grants No. 2016YFA0300401 and 2020YFA0308800), the National Natural Science Foundation of China (Grants No. 11874206, 12061131001, 92065109, 11734003 and 11904294), the Fundamental Research Funds for the Central Universities (Grant No. 020414380095), Jiangsu shuangchuang program, the Beijing Natural Science Foundation (Grant No. Z190006 and Z210006), and the Beijing Institute of Technology Research Fund Program for Young Scholars (Grant No. 3180012222011).
\end{acknowledgments}

%%%%%%%%%%%%%%%%%%%%%%%%%%%%%%%%%%%%%%%%%%%%%%%%%%%%%%%%%%%%%%%%%%%%%%%%%%%%%%%
% The bibliography (BibTeX)
%

\end{document}